# Validation of electrodeposited $^{241}$Am alpha-particle sources for use in liquified gas detectors at cryogenic temperatures

E. Calvo Alamillo [*], M.T. Crespo Vázquez [**], P.F. Rato Mendes [***], R. Álvarez Garrote, J.I. Crespo Anadón, C. Cuesta, A. De la Torre Rojo, I. Gil-Botella, I. Martín Martín, M. Mejuto Mendieta, C. Palomares, L. Pérez Molina, J.A. Soto Otón, A. Verdugo de Osa

*Centro de Investigaciones, Energéticas, Medioambientales y Tecnológicas (CIEMAT), Avda. Complutense 40, 28040, Madrid, Spain*

ABSTRACT

This paper describes a procedure for the validation of alpha-particle sources (exempt unsealed sources) to be used in experimental setups with liquefied gases at cryogenic temperatures (down to −196 °C) and high vacuum. These setups are of interest for the development and characterization of neutrino and dark matter detectors based on liquid argon, among others. Due to the high purity requirements, the sources have to withstand high vacuum and cryogenic temperatures for extended periods. The validation procedure has been applied to $^{241}$Am sources produced by electrodeposition.

## 1. Introduction

In neutrinos and dark matter detection experiments, large volume detectors using liquefied gases, such as liquid argon, are used. Radioactive sources are widely used in these experiments for calibration (Baldini et al., 2006; Agnes et al., 2017; Abi et al., 2020a). The scintillation light in liquid argon and xenon is produced with wavelengths in the vacuum ultraviolet region (below 200 nm); so, the calibration of the photon-detector systems cannot use light sources such as Lasers or LED's and requires the use of radioactive sources for exciting the liquefied gas. Alpha sources provide accurate detector calibration because of the short range of this particle and its localized energy deposition close to the source (Baldini et al., 2006). Such sources must be used immersed in the liquefied gas. Therefore, they must be capable of withstanding the specific temperature and pressure conditions of cryogenic operation and, in addition, resisting high vacuum conditions during the cleaning of the experimental setup. Cleaning prior to operation under cryogenic conditions is necessary to maintain the purity of the liquefied gas within the necessary parameters to carry out the experiments. In order to use alpha sources under these conditions, their integrity must be guaranteed over time to ensure that there is no dispersion of radioactive material and, consequently, radioactive contamination of the detector medium.

Within the R&D&I activities carried out by the CIEMAT (Centro de Investigaciones Energéticas, Medioambientales y Tecnológicas) Group of Neutrinos in Reactors and Accelerators, the commitment acquired with the DUNE experiment (Deep Underground Neutrino Experiment) (Abi et al., 2020b) for the characterization of photon-detection system elements for the ProtoDUNE-SP detector (Abi et al., 2020c; Abed Abud et al., 2022) at the Neutrino Platform of CERN (European Organization for Nuclear Research) is an important task. For the calibration of these elements based on silicon photomultipliers (SiPMs), which are used in neutrino and dark matter detectors with cryogenic liquefied gases (Agnes et al., 2015), alpha sources of $^{241}$Am prepared by the Ionizing Radiations Metrology Laboratory (LMRI) of CIEMAT have been used. This paper deals with the procedure to validate the stability of these radioactive sources not only under cryogenic conditions but also under high vacuum conditions during cleaning operations, both for prolonged periods. To this aim, their possible degradation over time and/or the loss of radioactive material in this type of environment has been studied.

The $^{241}$Am sources used for validation have been prepared by the

E. Calvo Alamillo et al.

method of electrodeposition. Among the different sources preparation methods (Greene et al., 1972; Lally and Glover, 1984; Aggarwal, 2016), electrolytic deposition, commonly known as electrodeposition, is the most common technique for alpha-sources preparation. Sources prepared by this method consist of a radioactive deposit onto a metallic substrate, commonly stainless steel, which acts as cathode of the electrolytic cell. Since the very early years of the last century, sources of alpha-emitting nuclides have been prepared by electrodeposition from organic or aqueous electrolytes (Broda, 1950; Rodden, 1950) although aqueous electrolytes are preferred since electrodeposition from organic media requires high voltages and the use of organics. In both cases, the actinide elements ($^{241}$Am in this paper) are too electropositive to be reduced at the cathode surface as metals and the deposit consists of an insoluble compound of the element, commonly oxy-hydroxides (Beesley et al., 2009; Méndez et al., 2010; Crespo, 2012). In the next sections, the methods followed to prepare and validate the sources used in this paper, as well as the experimental setups of detector prototypes, will be described in detail.

## 2. Materials and methods

### 2.1. Source preparation

The most widespread methods for the electrodeposition of actinides use aqueous electrolytes containing sulphate ions, being the Hallstadius (1984) method, that uses a sulphuric acid electrolyte with sodium sulphate to prevent the adsorption of the low mass concentration of most actinides onto the electrodeposition cell wall, one the most extensively used methods of alpha-sources preparation in routine measurements. This method has been applied to prepare the $^{241}$Am sources that have been tested in the experimental setups for the characterization of the SiPMs used in cryogenic liquefied gas detectors in this work.

The procedure for the preparation of the sources is as follows: An aliquot (0.4–0.5 cm$^3$) of a LMRI solution of $^{241}$Am with an activity concentration of 134.4±1.0 Bq/g ($k$=1) was transferred into a 10 cm$^3$ glass beaker and evaporated to dryness. Then, 1 cm$^3$ of 0.3 M Na$_2$SO$_4$ was added. The resulting solution was evaporated until dry and the residue was dissolved in 0.3 cm$^3$ of concentrated H$_2$SO$_4$. About 3–4 cm$^3$ of distilled water were added and the solution was transferred into the electrodeposition cell. The glass beaker was washed 2 times with 3–4 cm$^3$ of water until a final volume in the cell of 10 cm$^3$ was obtained. Two drops of 0.1 % thymol blue indicator were added and the pH was adjusted to approximately 2.2 with 25 % NH$_4$OH. The electrodeposition cells were made of polypropylene and the water tightness was checked before each use. The bottom of the cell is the cathode and the backing of the deposit. It consists on a mirror-polished stainless steel disk, first cleaned with acetone and then washed with deionised water, of 25 mm diameter and 1 mm thickness. The anode was a 1 mm diameter platinum wire folded in the base into a spiral shape of approximately 1.6 cm$^2$ of area. As the geometry of the anode has an important influence on the homogeneity of the alpha sources, the spiral shape has been found to produce sources with good homogeneity (Klemencic and Benedik, 2010; Jobbágy et al., 2013). During electrodeposition, the cell was immersed in a cold water bath to slightly cool the electrolyte and avoid excessive evaporation. Other experimental conditions are current intensity (1.5 A), cathode to anode distance (10 mm) and electrodeposition time (60 min). Also, it is important to highlight that the obtained electrodeposition yield for $^{241}$Am was of the order of 100 %, similar to that found by other authors (Hallstadius, 1984; Krmpotić et al., 2017). In this sense, the role played by platinum atoms co-deposited on the cathode surface, and the subsequent increase in the hydroxyl ions concentration, is very important to explain the high yields obtained by the Hallstadius method to prepare actinide sources (Beesley et al., 2009; Méndez et al., 2010). During electrodeposition in this sulphate medium, platinum atoms are dissolved from the anode and, once in solution, deposited on the cathode in the form of a porous structure, thus increasing the concentration of hydroxyl ions and enabling precipitation of insoluble hydroxides

### 2.2. Autoradiography

The activity distribution or homogeneity of the alpha sources was checked before the experiments by obtaining a bitmap of the autoradiograph of each source using a phosphor imaging autoradiography system. A film sensitive to radiation type (Kodak "Imaging Screen-K″) was placed on the sources in the inside of a black box to avoid ambient ultraviolet and visible light. After the exposure time, the film was scanned using laser technology type Pharos FX and the information is transformed into a bitmap using the software "Quantity One", in which detailed distribution of the activity on the film is displayed after filtering out the ultraviolet background noise using an image processing program. Different exposure times are usually tested to obtain the optimum one for the activity of the sources.

### 2.3. Activity measurement of the $^{241}$Am sources

The alpha counting rate of the $^{241}$Am sources has been measured in a $2\pi$ grid ionization chamber (NUMELEC NU 14B) which also allows determining the alpha-particle spectra of the sources. At least five measurements of each source, the same counting time selected to obtain adequate counting statistics, have been completed. The area of each spectrum, including extrapolation to zero energy and background reduction, divided by the counting time is the counting rate of a measurement. The mean value of all the values obtained for each source is the counting rate $C$ of the source. The activity is then obtained as:

$$A = \frac{2C}{1+B}$$

in which $B$ is the backscattering factor obtained from Crawford (1949) that depends on the alpha energy of the radionuclide and the atomic weight of the disk or backing of the deposit. The correction for self-absorption has not been included since the prepared sources are thin. Uncertainty associated to $A$ was obtained from the uncertainties of $C$ and $B$, the latter being estimated as 0.2 x $B$.

### 2.4. Cryogenic experimental setups

Three experimental setups were used. These are based on the operating conditions to which the sources were subjected: cryogenic temperatures down to −196 °C (liquid nitrogen), different pressures and high vacuum conditions, reaching values of up to $10^{-4}$ Pa ($10^{-6}$ mbar).

1) Preliminary assessment: fast immersion tests of thermal stress

In order to carry out rapid cooling and heating tests, a cryogenic container of 2-L volume, filled with liquid nitrogen and including a source holder has been designed (Fig. 1).

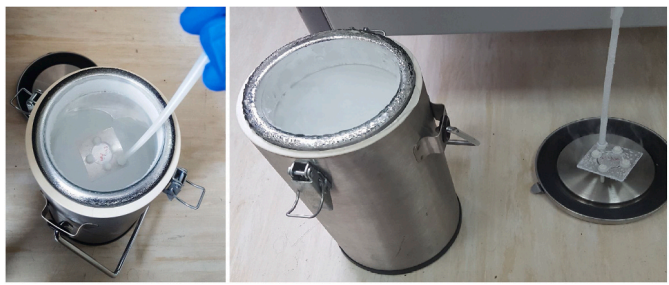

**Fig. 1.** Two-liter volume setup for fast immersion tests of thermal stress.

2) Long immersion tests under cryogenic temperatures and high vacuum. Three-liter setup.

Once the preliminary assessment validated the stability of the $^{241}$Am sources at the liquid nitrogen temperature, two experimental setups developed for the characterization of photodetectors using cryogenics and vacuum have been used in the next steps of the testing.

The first one is a small hermetic container of 3 L volume, inserted into a cryogenic container of 50 L (Fig. 2). This last container of liquid nitrogen allows liquefying argon (or nitrogen) introduced into the internal volume of 3 L, where the $^{241}$Am source is located, by cooling its walls. This is achieved by controlling pressure parameters. Automatic filling of the 50 L container with liquid nitrogen is carried out from the 450 L tank, which is at an overpressure of 5 bar, depending on the gas to be liquefied (nitrogen or argon). Before starting the filling and liquefying in the small 3 L container, vacuum is made for hours or days until reaching a value of around $10^{-4}$ Pa ($10^{-6}$ mbar). This is done to remove any air and humidity contained in the elements introduced in this container. The vacuum must be slowly reached to protect the detection elements located in this volume.

3) Long immersion tests under cryogenic temperatures and high vacuum. Eighteen-liter setup.

This second experimental setup consists of a 300 L external cryogenic container, which includes two internal concentric containers. The largest, where the liquid nitrogen is introduced, has a volume of about 100 L and within it a smaller one of about 18 L, where argon is liquefied and where the $^{241}$Am alpha source is located. The experimental configuration is the same as in the previous section, except for the volume of the cryogenic container.

## 3. Results and discussion

### 3.1. Americium-241 sources

Two $^{241}$Am sources were prepared by the electrodeposition technique (LMRI identifications FRC-2020-623 and FRC-2021-645). Their activity was determined before and after subjecting them to cryogenic temperatures and high vacuum, to determine any potential activity loss. A visual inspection of the sources showed no degradation after handling, which was confirmed by the spectral resolution obtained.

Fig. 3 shows autoradiographs of both sources at exposure times of 20, 45 and 270 min. Different exposure times were used to verify that there were no saturation effects at long exposures. The activity distribution shows that americium is distributed homogeneously in all cases, which confirms that the electrodeposition method on stainless steel disks with a spiral wound wire Pt anode produces good quality $^{241}$Am sources.

### 3.2. Validation procedure

The electrodeposited $^{241}$Am sources have been subjected to cryogenic temperatures down to $-196$ °C (liquid argon or liquid nitrogen conditions) at different pressures and high vacuum conditions of up to $10^{-4}$ Pa ($10^{-6}$ mbar). To validate their suitability to the characterization of neutrino detectors under these conditions, a two-phase validation procedure was designed.

The first phase (preliminary assessment) used the small and easy to decontaminate, if necessary, experimental setup for rapid tests of thermal stress, as described above. In this setup each source was subjected to rapid thermal stress of cooling and heating cycles, from liquid nitrogen to room temperatures. Ten cycles of 10 min cooling of the sources in liquid nitrogen, removal of the sources and 20 min heating at room temperature were applied. After these ten cycles, their activity was compared with the initial values to check possible losses of $^{241}$Am. No changes in the activity of the sources were detected (Table 1), and they were used for further testing in the next phase.

In the second phase, the behavior of the sources in the largest volume setups designed to operate in prolonged cooling cycles and high vacuum was studied. Once the source and the devices to be tested were introduced in the internal volumes of 3 L or 18 L of the experimental setups previously described, each cycle consisted of three steps: vacuum, cooling and heating. A high vacuum of approximately $10^{-4}$ Pa ($10^{-6}$ mbar) was obtained in these internal volumes with the use of a turbomolecular pump, taking from two days for the smallest internal volume to five days for the largest. Once the necessary vacuum was reached, the cooling cycle was started by filling the volume adjacent to the internal volume with liquid nitrogen. In this way, the internal volume begins to cool down slowly, therefore cooling the alpha source. Once a certain cryogenic temperature has been reached, the liquefaction process begins and the source will reach cryogenic equilibrium temperature. The system will continue to operate under these conditions for several days, maintaining temperature through the liquid nitrogen filling system and its automated pressure control. The cycle ends when liquid nitrogen is no longer injected and the temperature rises slowly until it reaches ambient conditions. This gradual heating process takes about one day for the 3 L internal volume setup and up to three days for the 18 L internal volume.

The FRC-2020-623 source has been subjected to three long cycles in the 3 L experimental setup and 10 cycles in the 18 L setup, which corresponds to about 30 days of high vacuum and about 77 days of cryogenics. Source FRC-2020-645 has only been subjected to 3 long cycles, tested in the 18 L setup, which corresponds to about 10 days of high vacuum and 27 days of cryogenics.

Table 1 shows the activity of both sources obtained after their preparation and after the two phases of the validation procedure. Taking into account the $^{241}$Am decay, all activity data have been expressed on the reference date of the first activity measurement.

## 4. Conclusions

In this paper we describe a procedure for the validation of low-activity electrodeposited $^{241}$Am alpha sources under high vacuum and cryogenic temperatures, for their use in liquefied gas detectors. We have produced and characterized two such sources and validated them according to this procedure. The results do not show any loss of activity of the sources since the activity values obtained after the different phases of the validation procedure have remained unchanged, within the measurement uncertainties. Therefore, it can be concluded that the source preparation method is suitable for the production of alpha-particle sources capable of withstanding the prolonged high vacuum periods and cryogenic temperatures required in high energy physics experiments using liquefied gas detectors.

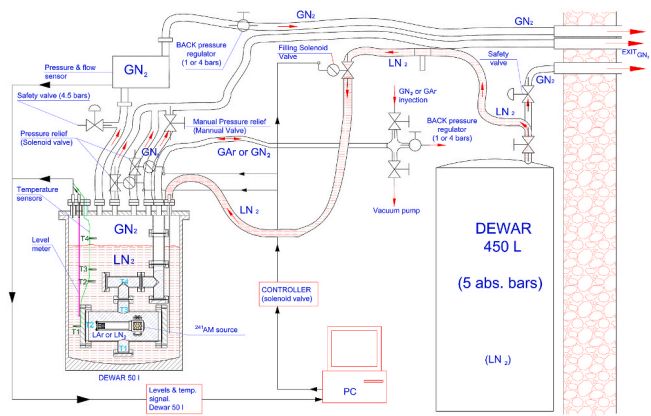

**Fig. 2.** Three-liter volume configuration scheme with hermetic container of 3 L volume inserted into a cryogenic container of 50 L with nitrogen that allows liquefying argon introduced into the internal volume of 3 L by cooling its walls. This is achieved by controlling pressure parameters.

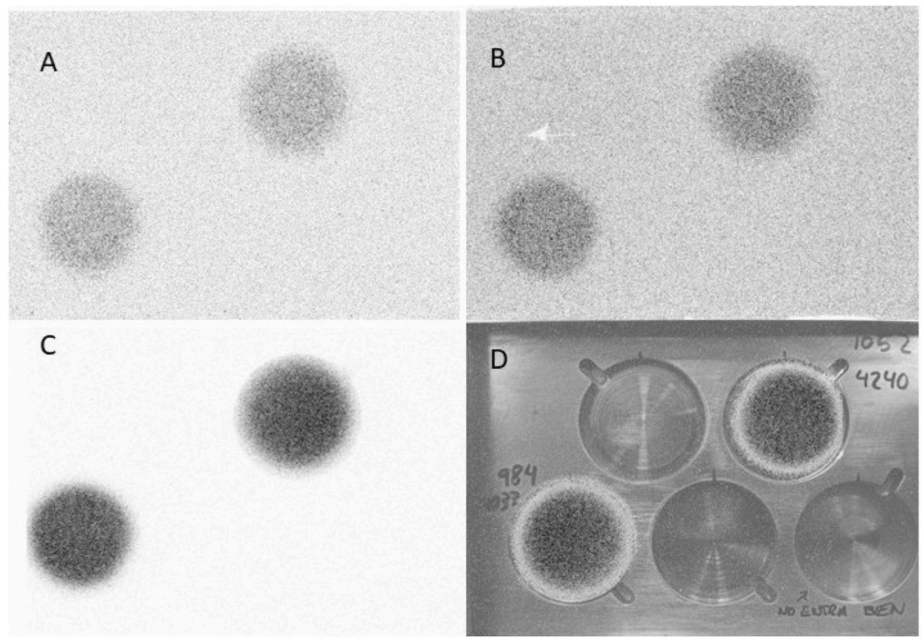

**Fig. 3.** Autoradiographs of each $^{241}$Am source with different exposure time: 20 min (A); 45 min (B); 270 min (C). Overlap of the 270 min exposure on the radioactive sources support (D).

**Table 1**
Activity of both sources obtained after their preparation and after the two phases of the validation procedure.

| SOURCE (Date of first activity measurement) | Measurements after preparation of the sources | Measurements after the first phase (preliminary assessment) | Measurements after the second phase (3 L experimental setup) | Measurements after the second phase (18 L experimental setup) |
|---|---|---|---|---|
| | Activity (Bq) | Activity (Bq) | Activity (Bq) | Activity (Bq) |
| FRC-2020-623 (12/08/2020) | 54,47 ± 0,27 | 54,55 ± 0,19 | 54,61 ± 0,22 | 54,66 ± 0,34 |
| FRC-2021-645 (08/03/2021) | 64,52 ± 0,33 | 64,23 ± 0,34 | – | 64,31 ± 0,24 |

### CRediT authorship contribution statement

**E. Calvo Alamillo:** Writing – review & editing, Writing – original draft, Supervision, Methodology, Investigation, Formal analysis, Data curation, Conceptualization. **M.T. Crespo Vázquez:** Writing – review & editing, Writing – original draft, Supervision, Methodology, Investigation, Formal analysis, Data curation, Conceptualization. **P.F. Rato Mendes:** Writing – review & editing, Writing – original draft, Supervision, Methodology, Investigation, Conceptualization. **R. Álvarez Garrote:** Formal analysis. **J.I. Crespo Anadón:** Formal analysis. **C. Cuesta:** Formal analysis. **A. De la Torre Rojo:** Software, Investigation, Formal analysis, Data curation. **I. Gil-Botella:** Supervision, Project administration, Methodology, Investigation, Conceptualization. **I. Martín Martín:** Software, Formal analysis, Data curation. **M. Mejuto Mendieta:** Data curation. **C. Palomares:** Investigation, Formal analysis, Data curation. **L. Pérez Molina:** Methodology, Investigation, Formal analysis, Data curation. **J.A. Soto Otón:** Formal analysis. **A. Verdugo de Osa:** Software, Methodology, Investigation, Formal analysis, Data curation, Conceptualization.

### Declaration of competing interest

The authors declare that they have no known competing financial interests or personal relationships that could have appeared to influence the work reported in this paper.

### Data availability

Data will be made available on request.

### Acknowledgments

This project has received funding from the European Union Horizon 2020 Research and Innovation programme under Grant Agreement no. 101004761; from Grant PID2019-104676 GB-C31 funded by MCIN/AEI/10.13039/501100011033; from the Comunidad de Madrid; and the support of a fellowship from "la Caixa" Foundation (ID 100010434) with code LCF/BQ/DI18/11660043.